\documentclass[%
reprint,
superscriptaddress,
 amsmath,amssymb,
 aps, prb,
]{revtex4-2}

\usepackage{gensymb}
\usepackage[dvipsnames]{xcolor}
\usepackage[colorlinks]{hyperref}
\hypersetup{
colorlinks=True,
linkbordercolor=White,
linkcolor=BrickRed,
citecolor=Blue,	 
}

\usepackage{siunitx}
\usepackage{graphicx}
\usepackage{layouts}
\usepackage{physics}
\usepackage{booktabs}

\usepackage[caption=false]{subfig}

\begin{document}
\vspace{-0.5cm}

\title{Gate-tunable magnetic damping in van der Waals Heterostructures}

\author{Lukas Cvitkovich}
\affiliation{Institute for Theoretical Physics, University of Regensburg, 93040 Regensburg, Germany}

\author{Klaus Zollner}
\affiliation{Institute for Theoretical Physics, University of Regensburg, 93040 Regensburg, Germany}

\author{Jaroslav Fabian}
\affiliation{Institute for Theoretical Physics, University of Regensburg, 93040 Regensburg, Germany}
\affiliation{\text{Halle-Berlin-Regensburg Cluster of Excellence CCE, University of Regensburg, 93040 Regensburg, Germany}}

\begin{abstract}
Active control of magnetic damping is a crucial capability for the development of low-power, tunable spintronic devices. In this work, we demonstrate that the intrinsic magnetic damping of 2D materials is highly sensitive to the position of the Fermi level relative to spin-orbit driven anti-crossings, or ``spin hot spots.'' Starting from a minimal multi-band model evaluated via Kamberský's breathing Fermi surface theory, we elucidate the fundamental role that energy gaps and band alignment play in governing magnetic dissipation. We then translate these theoretical insights to van der Waals heterostructures, proposing a concrete mechanism for strongly tunable magnetic damping via electrostatic gating. This concept is validated from first principles via density functional theory calculations on a realistic Fe$_3$GeTe$_2$/graphene heterostructure. Our calculations reveal that an out-of-plane electric field selectively shifts the relevant energy bands, enabling the modulation of the heterostructure's magnetic damping over several orders of magnitude. These findings establish a robust theoretical framework and a promising materials platform for electrically tunable magnetization dynamics.
\end{abstract}

\maketitle

The dynamic response and energy efficiency of spintronic devices~\cite{Zutic_RMP_2004} are fundamentally governed by magnetic damping~\cite{Gilbert_2004, Hickey_2009, Azzawi_2017}. While a low damping parameter minimizes the critical current required for spin-transfer torque switching, higher damping is necessary to ensure rapid magnetization settling times and robust thermal stability. Consequently, the ability to actively tune magnetic damping \textit{in operando} is a highly desirable milestone for adaptable spintronic architectures~\cite{Ralph_2008, Chiba_2021}. Conventional methods for tailoring damping, such as chemical alloying~\cite{Rantschler_2007, Schoen2016}, interfacing with heavy metals~\cite{Swindells_2022}, or twisting~\cite{Wang_twist_tune_damping_2025, Chen_2025} are inherently static and determined during fabrication. In traditional bulk metallic ferromagnets, Thomas-Fermi screening precludes direct electrostatic gating, limiting active electrical control to indirect methods such as strain-mediated phase transitions~\cite{Nan_2020}. However, the emergence of two-dimensional (2D) materials provides a pathway to completely bypass this electrostatic limit, establishing a uniquely tunable platform for next-generation devices.

At a microscopic level, the intrinsic magnetic damping in itinerant ferromagnets is dictated by the interplay between spin-orbit coupling (SOC) and the electronic band structure. Within the framework of Kamberský’s breathing Fermi surface (BFS) model~\cite{Kambersky_breathing_fermi_surface_1970, Kunes_2002}, a precessing magnetization dynamically deforms the Fermi surface, leading to the excitation of electron-hole pairs and subsequent energy dissipation. This dissipation is highly sensitive to the density of states (DOS) at the Fermi level $E_\mathrm{F}$~\cite{Qu_fermi_2014, Schoen2016}. Consequently, sweeping $E_\mathrm{F}$ represents a theoretically ideal mechanism for modulating macroscopic damping. However, accessing this tuning knob requires a material platform where carrier concentration and band alignment can be substantially altered \textit{in situ}.

Van der Waals (vdW) heterostructures~\cite{Geim_vdW_2013, Novoselov_2D_2016, Carr_twistronics_2017, Carr_electronic-structure_2020, Sierra_vdW_2021, Hennighausen_twistronics_2021, Ribeiro-Palau_twistable_2018, Kurebayashi2022} provide exactly this capability~\cite{Verzhbitskiy2020}. Among available 2D magnets~\cite{Gong2017, Huang_2dmag_2017, Xing_electric_field_2017, cvitkovich_ML_2025}, the metallic ferromagnet Fe$_3$GeTe$_2$ (FGT) has emerged as a premier candidate for spintronic applications due to its strong perpendicular magnetic anisotropy, high Curie temperature, robust metallicity, and proven susceptibility to electrostatic gating~\cite{Fei2018, Deng_gate_tunable_2018, Alghamdi2019,Barenfanger_2025}. By integrating FGT with other vdW materials, external gate voltages can be utilized not merely to inject charge, but to selectively shift the relative energy bands of the constituent layers. This gate-induced band alignment provides the precise physical mechanism needed to forcefully modulate the DOS of the magnetic layer at the Fermi level.

In this work, we demonstrate that this band-alignment mechanism can be exploited to achieve wide-range, active control of magnetic damping. We first validate the underlying physics using a minimal multi-band model, highlighting the extreme sensitivity of the damping parameter to the position of $E_\mathrm{F}$. We then quantify this effect from first principles using density functional theory (DFT) on a realistic FGT/graphene heterostructure. Because pristine graphene possesses inherently weak SOC~\cite{Gmitra_2009}, it contributes minimally to the total magnetic dissipation, serving as a necessary prerequisite to establish a low-damping regime. 
In addition, since graphene exhibits an exceptionally low DOS near the Dirac point, the applied electric field induces a massive shift in its chemical potential, whereas the FGT bands only shift by a few meV. Despite the FGT bands remaining largely pinned, our calculations reveal that the resulting change in relative band alignment modulates the heterostructure's overall magnetic damping by over two orders of magnitude.

\textit{Single-band model.} We start from a minimal single-band model which includes a mean-field exchange term as well as Rashba SOC
\begin{equation}
\label{eq:H_single}
\begin{split}
H &= H_\mathrm{kin}+ H_\mathrm{ex}+ H_\mathrm{SOC} \\ & = \frac{k^2}{2m} \sigma_0 + J (\mathbf{n} \cdot \boldsymbol{\sigma}) + \alpha_R (k_y \sigma_x - k_x \sigma_y),
\end{split}
\end{equation}
where $k$ is the magnitude of crystal momentum $\boldsymbol{k}=(k_x, k_y)$, $m$ is the effective mass, the magnetization direction $\mathbf{n}$ is defined by polar angle $\theta$ and azimuthal angle $\phi$ as $\mathbf{n} = (\sin\theta\cos\phi, \sin\theta\sin\phi, \cos\theta)$, $\alpha_\mathrm{R}$ and $J$ parametrize Rashba SOC and exchange, respectively, $\boldsymbol{\sigma}=(\sigma_x, \sigma_y, \sigma_z)$ are the Pauli matrices, $\sigma_0$ is the identity matrix, and $\hbar=1$. 
We then apply Kambersk{\'y}'s BFS model~\cite{Kambersky_breathing_fermi_surface_1970, Kunes_2002} to calculate the dimensionless damping tensor
\begin{equation}
\label{eq:BFS}
    \alpha_{i j} = \frac{g \mu_\mathrm{B}}{M \mu_0} \sum_n \int \frac{\mathrm{d}^2\boldsymbol{k}}{(2\pi)^2} \eta(E_{n,\boldsymbol{k}}) \frac{\partial E_{n,\boldsymbol{k}}}{\partial i}\frac{\partial E_{n,\boldsymbol{k}}}{\partial j} \frac{\tau}{\hbar},
\end{equation}
where $i,j\in \{\theta,\phi\}$, $g$ denotes the Landé $g$-factor, $\mu_\mathrm{B}$ is the Bohr magneton, $M$ is the total magnetization per unit volume, $\eta$ is the negative derivative of the Fermi function, $E_\mathrm{n,\boldsymbol{k}}$ are the energies labeled with wave vector $\boldsymbol{k}=(k_x,k_y)$ and band index $n$, and $\tau$ is the electron-hole lifetime (assumed to be equal for all bands in the following) generated by the precessing magnetization~\cite{Kambersky_breathing_fermi_surface_1970, Gilmore_2007}.

Diagonalization of the Hamiltonian and solving the integral Eq.~\eqref{eq:BFS} under the approximation that exchange dominates SOC ($J>>\alpha_\mathrm{R}k$) yields an analytical solution (see App.~\ref{app:BFS_analytic}) for the polar damping
\begin{equation}
\label{eq:damp_singleb}
    \alpha_{\theta\theta} = \frac{g \mu_\mathrm{B}}{2 M \mu_0} \tau \alpha_R^2 \cos^2\theta \sum_\mu D_\mu(E_\mathrm{F}) k_\mu^2
\end{equation}
where $D_\mu$ is the 2D density of states of spin band $\mu$ which acts as a step function because $D_\mu=\frac{m}{4\pi}$ if $E_F$ is above the band edge and zero otherwise, and $k_\mu^2=2m(E_\mathrm{F}\mp J)$ is the squared Fermi wave vector of each spin band which grows linearly with $E_F$.

\begin{figure}[htbp]
	\centerline{\includegraphics[width=\linewidth]{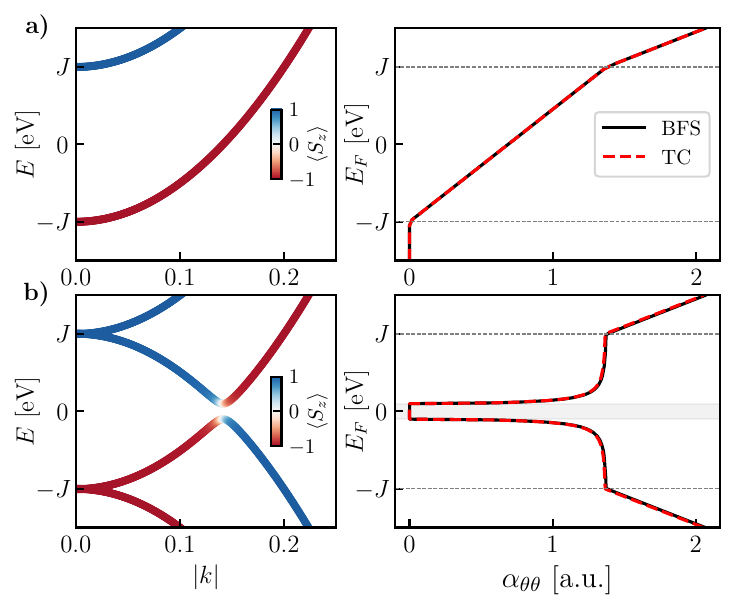}}
    \caption{Band structure and damping in the case of \textbf{a)} the single-band model and \textbf{b)} the two-band model.
    The left panels show the respective band structure color-coded by the spin expectation value $\langle S_z\rangle$. The right panels show the polar damping $\alpha$ versus $E_\mathrm{F}$ according to the BFS (black) and TC model (red).
    }
    \label{fig:bands_and_damping}
\end{figure}

The linear dependence of the magnetic damping on $E_\mathrm{F}$ is a result of three ingredients: the nature of the spin-orbit field, the dissipation mechanism, and the two-dimensional density of states. We consider a Rashba spin-orbit coupling (which acts as a perturbation proportional to momentum $k$) combined with a linear-response theory where the dissipated power scales with the square of the perturbation. Because the density of states for a 2D parabolic band is independent of energy, this $k^2$-dependency translates directly into a strictly linear dependence on $E_\mathrm{F}$ as the Fermi level sweeps past the respective band edges. 
The zero-temperature damping results for the model parameters $J=0.1$\,eV, $\alpha_R=0.01$, $m=0.1$\,eV$^{-1}$, $\tau=1$\,ps are shown in Fig.~\ref{fig:bands_and_damping}a~\footnote{To ensure numerical reproducibility, we note that the integration is performed over a dimensionless fractional k-space grid defined on the interval [-0.5, 0.5] × [-0.5, 0.5]. Consequently, to maintain a consistent energy scale, the Rashba strength $\alpha_R$ acts as a pure energy scale (eV) and the effective mass $m$ acts as an inverse energy scale (eV$^{-1}$). Because this phase-space integration is not normalized against a specific material's Brillouin zone area, atomic volume, or saturation magnetization, the calculated Gilbert damping represents a relative scaling factor and is thus reported in arbitrary units (a.u.).}. 
As shown in App.~\ref{app:BFS_analytic}, this result is also valid for the angular damping $\alpha_{\phi\phi}$.
For completeness, we confirm this result by means of the more sophisticated torque correlation (TC) model in Fig.~\ref{fig:bands_and_damping} (see also App.~\ref{app:TC}).

\textit{Two-band model.} In a next step, we extend the single-band model to a two-band model expressed in the basis $\{ |1, \uparrow\rangle, |1, \downarrow\rangle, |2, \uparrow\rangle, |2, \downarrow\rangle \}$
 by the Hamiltonian
\begin{equation}
\label{eq:twobandH}
H =
\begin{pmatrix}
\epsilon_1 + J_z & \Gamma & 0 & V_\mathrm{SO} \\
\Gamma^{*} & \epsilon_1 - J_z & V_\mathrm{SO} & 0 \\
0 & V_\mathrm{SO}^{*} & \epsilon_2 + J_z & \Gamma \\
V_\mathrm{SO}^{*} & 0 & \Gamma^{*} & \epsilon_2 - J_z
\end{pmatrix},
\end{equation}
where $\Gamma = J_- + i\alpha_\mathrm{R} k_-$ with $J_{\pm} = J_x \pm iJ_y$ and $k_{\pm} = k_x \pm ik_y$. 
The top-left and bottom-right $2\times2$ blocks share the same form as the single-band Hamiltonian in Eq.~\eqref{eq:H_single}, but are allowed to have different effective masses such that $\epsilon_i=E_i+\frac{k^2}{2m_i}$ for $i=\{1,2\}$. To enable an interaction between the two bands, we introduce an inter-band spin-orbit coupling  $V_\mathrm{SO}$.

The resulting band structure is presented alongside a numerical solution for the BFS and TC damping within the two-band model in Fig.~\ref{fig:bands_and_damping}b. The parameters are set equal to the model of Fig.~\ref{fig:bands_and_damping}a; additionally, $V_\mathrm{SO}=0.01$\,eV, and the masses of the two bands have equal magnitude as before but opposite sign $m_1 = -m_2 = m$.
If $V_\mathrm{SO}=0$, the two-band model contains essentially the same physics as its single-band predecessor, however, with a second set of mirrored bands. Similarly, the total damping as a function of $E_\mathrm{F}$ would be the sum of linear contributions from all four spin-resolved bands -- a simple extension of Eq.~\eqref{eq:damp_singleb} -- leading to a constant damping for $|E_\mathrm{F}|<J$.

At finite coupling $V_\mathrm{SO}$, states of opposite spin mix, open up a gap (here at $E=0$), and form what is referred to as a ``spin hot spot''~\cite{Fabian1998, Raith2012}. This band topology manifests in the damping as a sharp drop to zero within the gap $|E_\mathrm{F}|<V_\mathrm{SO}$. Close to the band edge, the damping is extremely sensitive to the position of the Fermi level. This opens a pathway for tunable magnetic damping by means of dynamical band structure engineering. 

\textit{Application in a vdW heterostructure.} Van der Waals (vdW) heterostructures offer a versatile platform for tuning magnetic damping via external electric fields. By shifting the energy bands of adjacent layers relative to one another, an applied field drives charge transfer between the layers. Crucially, this process sweeps the Fermi level through the band structure, altering the effective density of states (DOS) at the Fermi energy. By integrating a low-SOC material like graphene with a 2D ferromagnet in a vdW heterostruture, the intrinsic disparity in their damping capabilities can be actively exploited to switch between macroscopic states of high and low damping.

As demonstrated above, this tuning works specifically well if the Fermi level lies close to an spin-orbit driven anti-crossing, or spin hot spot. 
In a next step, we go beyond these qualitative models and employ a more realistic approach for obtaining quantitative predictions, namely density functional theory (DFT).
To this end, we choose the material system Fe$_3$GeTe$_2$/graphene as a representative example.

\textit{Material properties.}
In the FGT/graphene heterostructure, we find that graphene is $n$-doped with a charge density of about $11.5\times 10^{12}$\,cm$^2$ at zero field.
Graphene's inherently weak SOC means it contributes minimally to the total magnetic damping of the heterostructure compared to the highly dissipative FGT layer. This heavy contrast is a key prerequisite for establishing a low-damping state in the system.
However, utilizing pristine graphene introduces a distinct challenge: when an out-of-plane electric field is applied, the graphene energy bands shift dramatically while leaving the FGT bands largely unaffected. This occurs because the shift in the chemical potential $\Delta\mu$, which dictates the band alignment relative to $E_\mathrm{F}$, is inversely proportional to the DOS ($\Delta\mu\propto 1/D$). Because pristine graphene has a low DOS near the Dirac point, it absorbs a large fraction of the applied potential difference. We show the band structures with projections on FGT and graphene for different fields in App.~\ref{app:bs}.
Surprisingly, even though the bands of the FGT are comparatively shielded and only weakly susceptible to the external field, we observe that the magnetic damping of the heterostructure can be tuned over more than two orders of magnitude~\footnote{Although this wide range is facilitated by large electric fields which would be challenging to apply in practice}, as discussed below.

\begin{figure}[htbp]
	\centerline{\includegraphics[width=\linewidth]{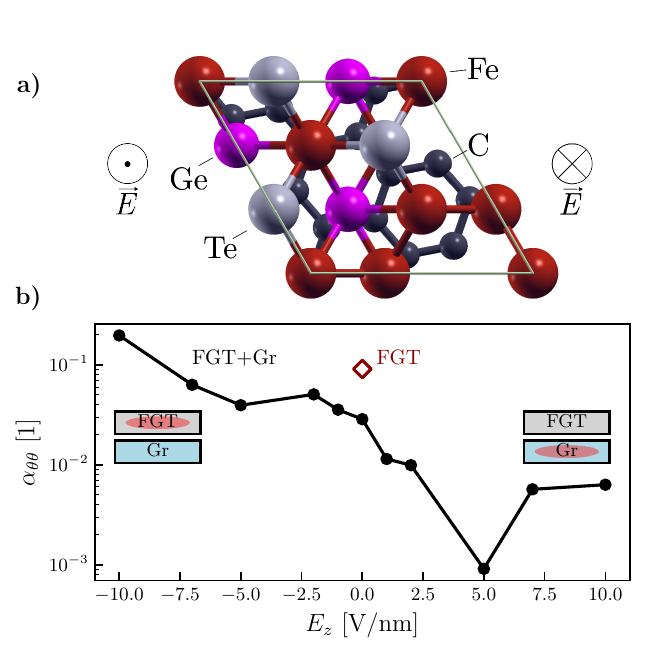}}
    \caption{Field-controlled polar damping in FGT/Gr for magnetization along $z$. \textbf{a)} Atomic model 
    of the  commensurate supercell used for DFT calculations. The vector direction symbols $\odot$ and $\otimes$ correspond to the negative and positive field regimes, respectively. 
    \textbf{b)}~Damping according to the BFS model versus out-of-plane electric field. In the schematic heterostructure, the ellipse represents charge that is transferred to the respective layer by the electric field.}
    \label{fig:DFT}
\end{figure}

\textit{DFT simulations.} We consider a prototypical vdW heterostructure composed of graphene and FGT with a twist angle of 10.9\degree, wherein the graphene layer is strained by 4.9\% in order to form a commensurate simulation cell with the relaxed FGT (lattice parameter 3.942\,\AA), see Fig.~\ref{fig:DFT}a~\footnote{Here, we conduct a proof of principle study and not a full scan upon the parameter space of twist, strain, interlayer distance, etc.}. The interlayer distance is 3.62\,\AA, measured from the graphene layer to the directly neighboring Te atoms of FGT. In total, there are 32 atoms in the supercell, 14 of which are C atoms. We utilize the plane-wave DFT code \textit{Quantum Espresso}~\cite{Giannozzi_2009}. Further details of the DFT setup are given in App.~\ref{app:DFT}. 
The damping is evaluated using the BFS model based on the DFT band energies. Because the BFS and TC formulations yield equivalent results for the band structure described by our model Hamiltonian, the computationally efficient BFS approach fully captures the relevant intra-band relaxation dynamics without loss of accuracy. 

Based on the DFT band energies, the damping is evaluated using the BFS model. This approach is justified by the results of our model Hamiltonian (Fig.~\ref{fig:bands_and_damping}), which demonstrate that the BFS and TC formulations yield equivalent results for such band structures. Furthermore, as established in Ref.~\cite{Gilmore_2007}, the full TC model naturally reduces to the BFS model in the limit of large electron relaxation times $\tau$. 


The results in Fig.~\ref{fig:DFT}b show that the damping in the heterostructure is already suppressed in comparison to a FGT monolayer ($\alpha_{\theta\theta}=0.029$ vs. $0.09$)~\footnote{These results are obtained by removing the graphene layer from the DFT supercell, leaving the FGT relatively unaffected to allow for the comparison.}. However, by applying an external field towards the FGT layer, the damping increases strongly (up to $\alpha_{\theta\theta}=0.20$), while it drops by another order of magnitude (down to $\alpha_{\theta\theta}=9\times10^{-4}$) if the field points towards the graphene layer. Considering experimentally accessible electric fields of $\pm2$\,V/nm, the magnetic damping can be modulated by a factor of five (from 0.05 to 0.01). See App.~\ref{app:table} for a complete list.

The decrease with increasing field is not monotonous since there are several bands whose damping contributions alter as the field varies.
Nevertheless, in the following, we show that the overall damping behavior can indeed be explained by the position of bands and spin hot spots with respect to the Fermi level, analogous to the simple considerations of Fig.~\ref{fig:bands_and_damping}b.

\begin{figure}[htbp]
	\centerline{\includegraphics[width=\linewidth]{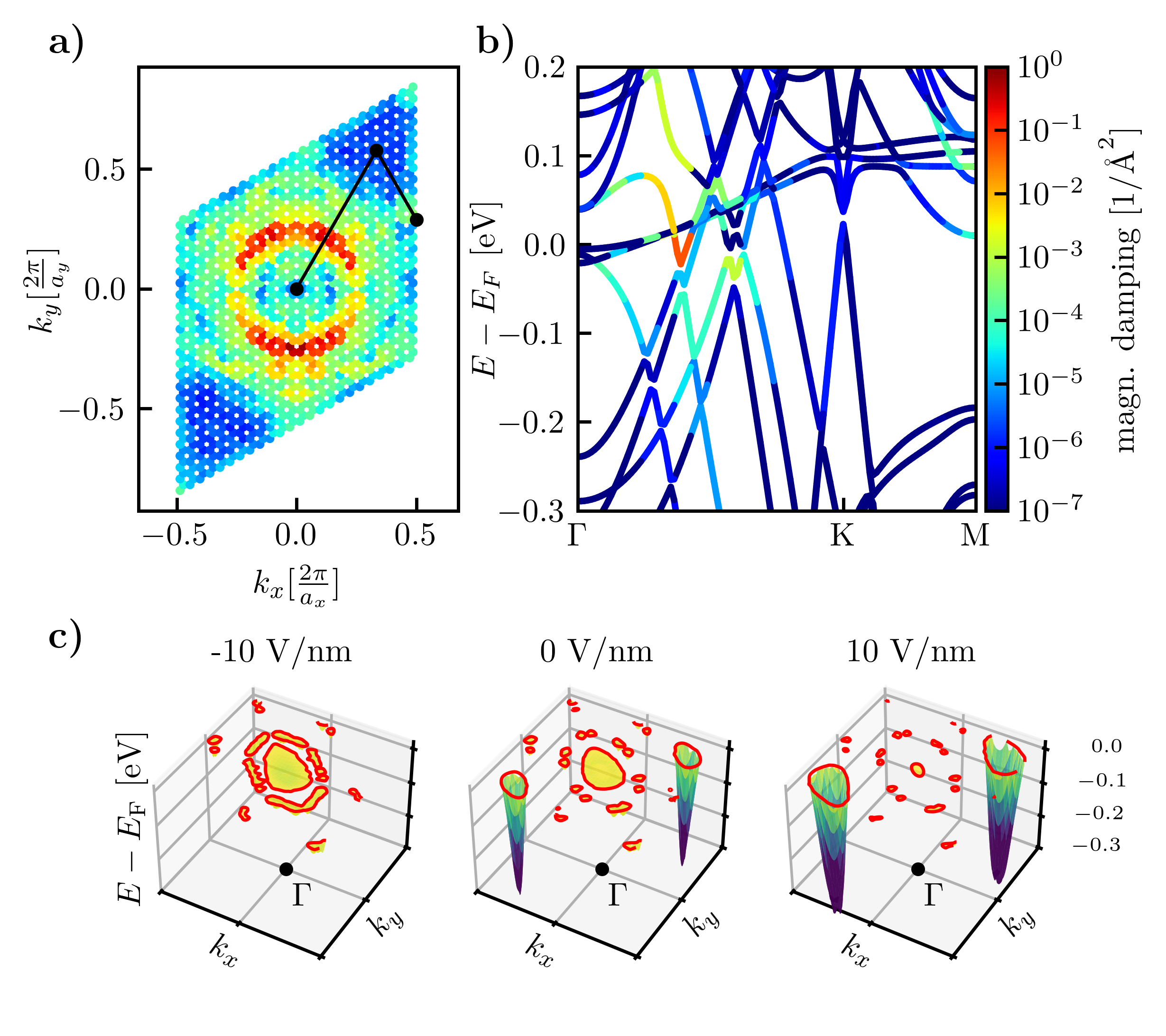}}
    \caption{Band- and $k$-resolved damping. \textbf{a)} Contributions to the damping across the Brillouin zone at $E_z=-10$\,V/nm. Damping is mainly generated in the red ring around $\boldsymbol{k}=0$. \textbf{b)} Band structure along the $k$-path indicated by the black line in a). Main driver of the damping is a single band associated with the ring in a). \textbf{c)} 3D plot and Fermi surface (red line) of this specific band. The band minimum can be shifted above (below) the Fermi level by the electric field which diminishes (enhances) the damping.
    }
    \label{fig:kspace}
\end{figure}

\textit{Contributions in $k$-space.} In order to reveal the dominant damping mechanism, we plot the $k$-resolved damping in Fig.~\ref{fig:kspace}a. The damping is mostly generated in a relatively narrow, circular region around the $\Gamma$-point at $\boldsymbol{k}=0$, so in the FGT bands (cf.~App.~\ref{app:bs}). Notably, hybridization with the graphene $p_z$-orbitals does not alter the damping behavior substantially, as regions around K, where the graphene Dirac cone lives, show virtually no damping.

Further insight is provided by the band- and $k$-resolved damping in Fig.~\ref{fig:kspace}b, which shows that the damping is dominated by a single hybridized band corresponding to the ring in Fig.~\ref{fig:kspace}a. This band exhibits a sharp minimum at finite $k$ driven by a spin hot spot -- more specifically, an anti-crossing between two steeply dispersing bands of opposite spin. It is precisely this anti-crossing behavior that allows the system to qualitatively realize the scenario described by the model Hamiltonian in Eq.~\eqref{eq:twobandH}. Because this band minimum resides just below the Fermi level, the system is perfectly positioned for efficient tuning. As demonstrated by our model, shifting $E_\mathrm{F}$ directly into the anti-crossing gap dramatically suppresses the magnetic damping.

Figure~\ref{fig:kspace}c demonstrates the shift of the relevant energy band by an out-of-plane electric field. Truncating the band at the Fermi energy visualizes the corresponding Fermi surface. Most importantly, this band forms shallow electron pockets that form a ring around $\boldsymbol{k}=0$. Because the band minima lies only about 0.02\,eV below the Fermi level (at zero field), it can be shifted above and below the Fermi level by the electric field. Thus, the damping generated in this band becomes extremely sensitive to the electric field.
At $+10$\,V/nm, the band lies almost completely above $E_\mathrm{F}$ such that the total damping in the heterostructure is strongly diminished.
Simultaneously, the Dirac cone is shifted down by the electric field. As discussed above, the shift of bands of the low-DOS layer is much larger compared to the large-DOS material. Thereby, the Dirac cone increases its Fermi surface drastically. However, this does not compensate for the overall reduction of the damping constant, due to the weak SOC in graphene.

\begin{figure}[htbp]
	\centerline{\includegraphics[width=\linewidth]{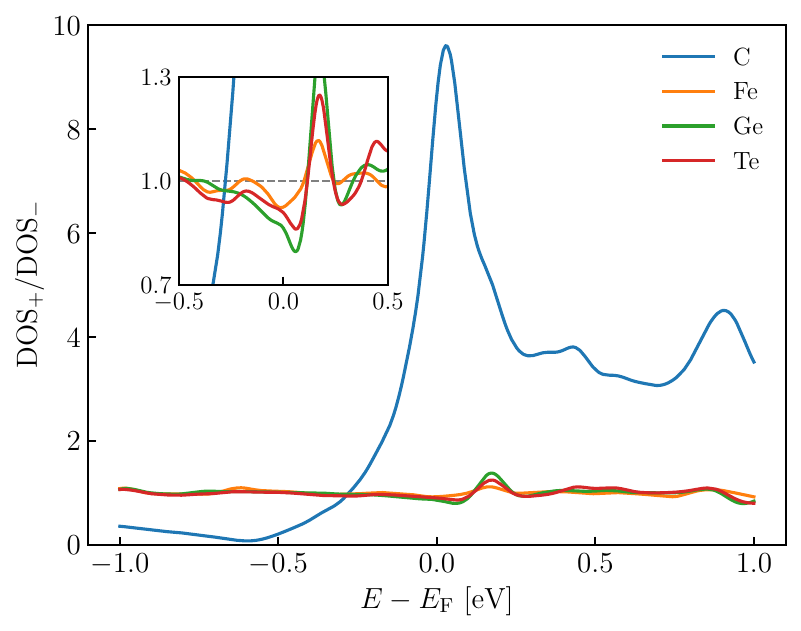}}
    \caption{Ratio of the atom-projected density of states (DOS) under reversed electric fields, DOS$_+$/DOS$_-$, where the subscripts $+$ and $-$ denote $E_z = +10$\,V/nm and $-10$\,V/nm, respectively. At the Fermi level, the ratio highlights a tenfold increase in the graphene DOS, contrasted with a minor $\sim$10\% reduction in the FGT layer DOS. The inset details the slight depletion of FGT states at $E_\mathrm{F}$ and their corresponding shift to higher energies ($\approx$\,0.15\,eV above $E_\mathrm{F}$).}
    \label{fig:DOS}
\end{figure}

\textit{Density of states.} To further elucidate the physical mechanism underlying the tunable damping, we analyze the response of the density of states (DOS) to the applied electric field (Fig.~\ref{fig:DOS}). The pronounced energy shift of the Dirac cone leads to a substantial accumulation of graphene states at the Fermi level. Specifically, the graphene DOS at $E_\mathrm{F}$ increases by a factor of 10 when the field is reversed from $E_z=-10$\,V/nm to +10\,V/nm. In contrast, the FGT DOS exhibits a minor reduction of approximately 10\,\% at the Fermi level. As illustrated in the inset of Fig.~\ref{fig:DOS}, this slight depletion occurs because the FGT states are shifted to higher energies, evidenced by a localized increase in the DOS roughly 0.15\,eV above $E_\mathrm{F}$.

As mentioned above, for a given induced charge density, the shift in the chemical potential is inversely proportional to the DOS. In the present material system, relatively large electric fields are required to induce FGT band shifts of a few tenths of an eV because a significant portion of the field is consumed by shifting the graphene states. 
Replacing graphene with a material that exhibits a larger DOS at the Fermi level would reduce this bottleneck, thereby lowering the required range of electric fields needed to achieve the same degree of damping.
Ultimately, the optimal design for achieving high electric-field tunability in vdW heterostructures involves pairing a magnetic layer with a material that exhibits both a high DOS and low intrinsic damping.

\textit{Conclusions.} Our findings establish electrostatic gating as a powerful tool for actively tuning magnetic damping in 2D materials. By shifting the Fermi level in the vicinity of spin hot spots, the damping can be modulated dramatically. We validated this mechanism first through a minimal multi-band model and subsequently via first-principles calculations on a realistic FGT/graphene van der Waals heterostructure. Crucially, by pairing a high-damping ferromagnet like FGT with a low-damping material like graphene, the intrinsic disparity in their properties can be exploited. Our calculations demonstrate that the magnetic damping of the heterostructure can be reliably controlled over at least one order of magnitude, providing a concrete blueprint for designing tunable spintronic devices.

This project has received funding from the Deutsche Forschungsgemeinschaft (DFG, German Research Foundation) 
SPP 2244 (Project No. 443416183), SFB 1277 (Project. No. 314695032), and the EU 2DSPIN-TECH Program (Graphene Flagship).

\bibliography{my.bib}

\clearpage

\appendix
\section{Derivation of Eq.~\eqref{eq:damp_singleb}}
\label{app:BFS_analytic}

\subsection{Effective Hamiltonian and Eigenvalues}

The complete Hamiltonian is defined by:
\begin{equation}
    H = H_{\text{kin}} + H_{\text{ex}} + H_{\text{SOC}}
\end{equation}
\begin{equation}
    H = \frac{k^2}{2m} \sigma_0 + J (\mathbf{n} \cdot \boldsymbol{\sigma}) + \alpha_R (k_y \sigma_x - k_x \sigma_y),
\end{equation}
where the magnetization direction $\mathbf{n}$ is parameterized as $\mathbf{n} = (\sin\theta\cos\phi, \sin\theta\sin\phi, \cos\theta)$. We group the exchange and Spin-Orbit Coupling (SOC) terms into an effective magnetic field $\mathbf{B}_{\text{eff}}$:
\begin{equation}
    H = \frac{k^2}{2m} \sigma_0 + \mathbf{B}_{\text{eff}} \cdot \boldsymbol{\sigma}
\end{equation}
\begin{equation}
    \mathbf{B}_{\text{eff}} = \left( J\sin\theta\cos\phi + \alpha_R k_y, J\sin\theta\sin\phi - \alpha_R k_x, J\cos\theta \right)
\end{equation}

The eigenvalues of this $2 \times 2$ matrix are:
\begin{equation}
    E_{\pm}(\boldsymbol{k}) = \frac{k^2}{2m} \pm |\mathbf{B}_{\text{eff}}|
\end{equation}

Applying the approximation that $J \gg \alpha_R k$, we expand the effective field magnitude:
\begin{align}
    |\mathbf{B}_{\text{eff}}| &= \sqrt{ J^2 + 2J\alpha_R (k_y \sin\theta\cos\phi - k_x \sin\theta\sin\phi) + \alpha_R^2 k^2 } \\
    &\approx J + \alpha_R \sin\theta (k_y \cos\phi - k_x \sin\phi)
\end{align}

Switching momentum into polar coordinates $\boldsymbol{k} = (k, \gamma)$, we substitute $k_y \cos\phi - k_x \sin\phi = k \sin(\gamma - \phi)$, yielding the approximated bands:
\begin{equation}
\label{eq:diag_energies}
    E_{\pm}(\boldsymbol{k}) \approx \frac{k^2}{2m} \pm J \pm \alpha_R k \sin\theta \sin(\gamma - \phi)
\end{equation}

\subsection{Breathing Fermi Surface Damping Formulation}

Within the framework of the breathing Fermi surface (BFS) model for a 2D system, the damping tensor elements $\alpha_{ij}$ (for $i, j \in \{\theta, \phi\}$) are given by:
\begin{equation}
    \alpha_{ij} = C \sum_{\mu = \pm} \int \frac{d^2k}{(2\pi)^2} \tau_{\mu} \left( \frac{\partial E_{\mu}}{\partial i} \right) \left( \frac{\partial E_{\mu}}{\partial j} \right) \delta(E_\mathrm{F} - E_{\mu})
\end{equation}
where $C=\frac{g \mu_\mathrm{B}}{M \mu_0}$ and $\tau_{\mu}$ is the electron-hole pair lifetime in band $\mu$. The required energy derivatives are:
\begin{align}
    \frac{\partial E_{\pm}}{\partial \theta} &\approx \pm \alpha_R k \cos\theta \sin(\gamma - \phi) \\
    \frac{\partial E_{\pm}}{\partial \phi} &\approx \mp \alpha_R k \sin\theta \cos(\gamma - \phi)
\end{align}

\subsection{Analytic Solution for Tensor Components}

The tensor is diagonal because the cross term $\alpha_{\theta\phi}$ depends on $\langle \sin(\gamma-\phi)\cos(\gamma-\phi) \rangle = 0$. 

Evaluating the radial integral over the delta function directly yields the spin-resolved density of states (DOS) at the Fermi level, $D_\mu(E_\mathrm{F}) = \int_{0}^{\infty} \frac{k dk}{2\pi} \delta(E_\mathrm{F} - E_\mu)$. For a 2D parabolic band, the DOS naturally behaves as a step function: $D_\mu = m/2\pi$ if the band is occupied ($E_\mathrm{F} > \pm J$), and $D_\mu = 0$ if it is empty. 

\paragraph{Polar Damping ($\alpha_{\theta\theta}$):}
\begin{equation}
\begin{split}
    \alpha_{\theta\theta} &= C \sum_{\mu = \pm} \tau_{\mu} \int_{0}^{\infty} \frac{k dk}{2\pi} \int_{0}^{2\pi} \frac{d\gamma}{2\pi} \left( \pm \alpha_R k \cos\theta \sin(\gamma - \phi) \right)^2 \\ 
    &\quad \times \delta(E_\mathrm{F} - E_{\mu})
\end{split}
\end{equation}

The angular integral evaluates to $1/2$. Substituting the DOS $D_\mu(E_\mathrm{F})$ isolates the squared Fermi momentum $k_{\mu}^2 = 2m(E_\mathrm{F} \mp J)$, yielding:
\begin{equation}
    \alpha_{\theta\theta} = \frac{C}{2} \alpha_R^2 \cos^2\theta \sum_{\mu = \pm} \tau_{\mu} D_\mu(E_\mathrm{F}) k_{\mu}^2
\end{equation}

\paragraph{Azimuthal Damping ($\alpha_{\phi\phi}$):}
By an identical procedure, using the angular average $\langle \cos^2(\gamma - \phi) \rangle = 1/2$, the azimuthal damping is:
\begin{equation}
    \alpha_{\phi\phi} = \frac{C}{2} \alpha_R^2 \sin^2\theta \sum_{\mu = \pm} \tau_{\mu} D_\mu(E_\mathrm{F}) k_{\mu}^2
\end{equation}

\begin{figure}[hbpt]
	\centerline{\includegraphics[width=\linewidth]{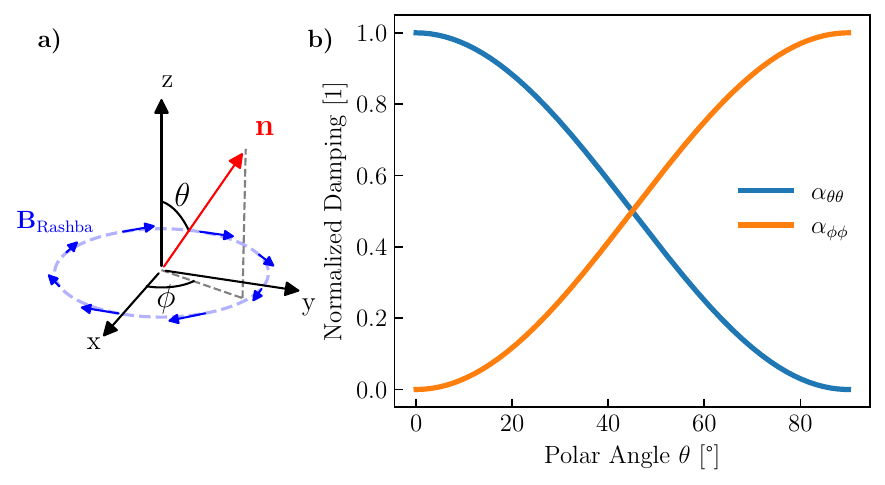}}
    \caption{
    \textbf{(a)} Schematic representation of the system geometry and effective magnetic fields. The macroscopic magnetization vector $\boldsymbol{n}$ is defined by the polar angle $\theta$ and azimuthal angle $\phi$.
    \textbf{(b)} The two components of the BFS damping tensor normalized to their maximum as a function of polar angle $\theta$.
    }
    \label{fig:damping_angle}
\end{figure}

The $\theta$-dependence of the tensor elements is plotted in Fig.~\ref{fig:damping_angle}. The polar damping reaches its maximum at $\theta=0$, that is for out-of-plane magnetization, and a minimum for $\theta=90$°. The azimuthal damping displays a complementary behavior, peaking at $\theta=90°$. We interpret these results by the following simple considerations.

Within the BFS model, the anisotropic magnetic damping arises from the variation of single-particle eigenenergies as the macroscopic magnetization direction $\boldsymbol{n}$ evolves. This ``breathing'' of the Fermi surface is governed by the relative alignment between the exchange field and the local spin-orbit field. 
Because the structural inversion asymmetry confines the Rashba effective field $\mathbf{B}_\mathrm{Rashba}$ strictly to the two-dimensional plane, only the in-plane projection of the magnetization ($\propto \sin\theta$) contributes to the first-order Zeeman energy shifts, see Eq.~\eqref{eq:diag_energies}. Consequently, the rate of change of these energies with respect to a polar tilt is maximized when the magnetization is nearly out-of-plane, leading to a polar damping component $\alpha_{\theta\theta} \propto \cos^2\theta$. Conversely, when the magnetization lies purely in-plane ($\theta = 90^\circ$), the in-plane projection is at a local maximum; small polar tilts produce no first-order energy shifts, causing the polar damping to strictly vanish.
Conversely, dynamic variations in the azimuthal angle $\phi$ modulate the alignment between the in-plane magnetization component and the local, spin-momentum locked Rashba fields of individual electron states. This in-plane rotation maximizes the shifting of the subband energies, resulting in continuous intraband scattering and an azimuthal damping $\alpha_{\phi\phi} \propto \sin^2\theta$ that peaks when the magnetization lies entirely within the two-dimensional plane.


\section{The torque correlation model}
\label{app:TC}

Compared to other proposed damping mechanisms~\cite{BlandHeinrich_2005} such as the BFS model, the torque-correlation (TC) model of Kambersk{\'y}~\cite{Kambersky_torque_correlation_1976} stands out due to its quantitative agreement with experiments. In contrast to the BFS model, it also considers scattering between two different bands~\cite{Gilmore_2007} and predicts
\begin{equation}
\label{eq:TC}
\alpha_{ij} = \frac{g \mu_\mathrm{B}}{M \mu_0} \sum_{n,m} \int \frac{d^{3}\boldsymbol{k}^2}{(2\pi)^{2}} \Gamma^{i}_{nm}(\boldsymbol{k})\Gamma^{j *}_{nm}(\boldsymbol{k}) W_{nm}(\boldsymbol{k}),
\end{equation}
where the matrix elements $\Gamma^{i}_{nm}(\boldsymbol{k}) = \langle n, \boldsymbol{k}|  [\sigma^i, H_{SO}]|m, \boldsymbol{k} \rangle$ measure spin-orbit-torque induced transitions between states $|n,\boldsymbol{k}\rangle$ and $| m, \boldsymbol{k}\rangle$. Here, $\sigma^i$ represents the Pauli spin operator projected along the local transverse spatial direction $i$ (i.e., $\sigma^\theta=\boldsymbol{\sigma}\cdot\hat{e}_\theta$ and $\sigma^\phi=\boldsymbol{\sigma}\cdot\hat{e}_\phi$). 
As shown in Ref.~\cite{Gilmore_2007}, the BFS and TC model are equivalent in the limit of low scattering.

\section{DFT setup}
\label{app:DFT}
We utilize the plane-wave DFT code \textit{Quantum Espresso}~\cite{Giannozzi_2009} to evaluate the band energies within a set of non-collinear calculations. The exchange-correlation interactions are treated within the generalized gradient approximation (GGA) using the Perdew-Burke-Ernzerhof (PBE)~\cite{PBE} parametrization in conjunction with the fully relativistic ultrasoft pseudopotentials~\cite{Vanderbilt_soft} based on the Rappe-Rabe-Kaxiras-Joannopoulos (RRKJ) optimization scheme~\cite{RRKJ_1990}. Furthermore, we used Grimme's DFT-D3 vdW correction~\cite{Grimme_2010}.
The plane-wave kinetic energy cutoffs for expanding the wavefunctions and the charge density are set to 120\,Ry and 960\,Ry, respectively. For the Brillouin zone integration, a dense $30\times30$ Monkhorst-Pack $k$-point grid is employed, and the electronic occupancies are treated using the Marzari-Vanderbilt smearing method with a broadening parameter of 0.005\,Ry. The self-consistent field (SCF) electronic minimization is converged to a threshold of $1\times10^{-12}$\,Ry.

For the evaluation of the BFS model, the derivative of the energies with respect to the magnetization angle is calculated numerically. We therefore conduct two calculations for each electric field strength. In the first calculation, the magnetization is not constrained in any way, and -- as expected -- relaxes to a ground state with out-of-plane magnetization. In the second calculation, the polar angle of the magnetization is tilted by an angle of 0.1° while the azimuthal angle is still unconstrained.

\section{Projected band structures}
\label{app:bs}

\begin{figure*}[htbp]
	\centerline{\includegraphics[width=0.8\linewidth]{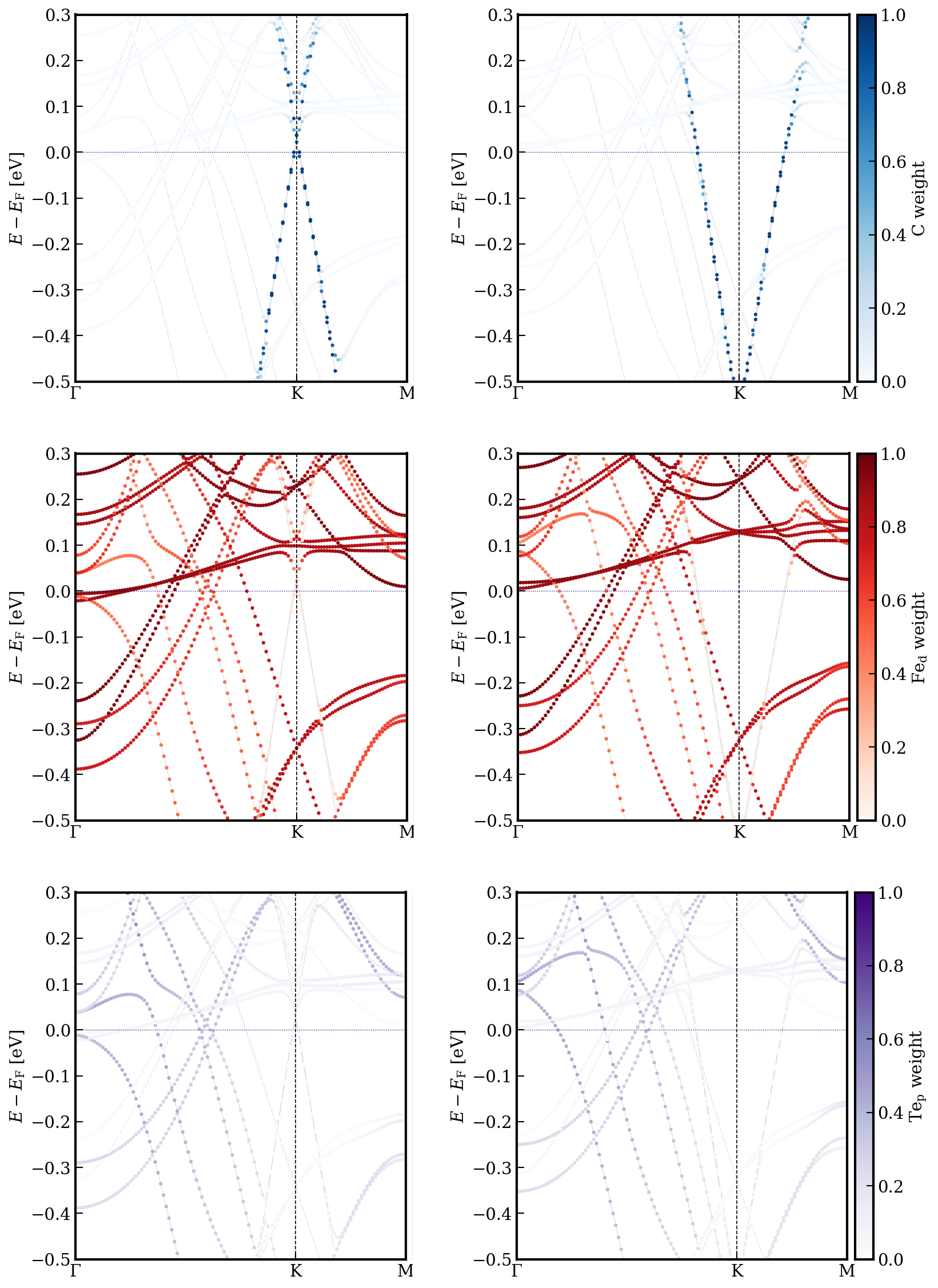}}
    \caption{Band structure of the FGT/Gr heterostructure for $E_z=-10$\,V/nm (left) and $E_z=+10$\,V/nm (right). The color map represents the fractional contribution (spectral weight) of all C orbitals (top), Fe$_\mathrm{d}$ orbitals (center), and Te$_\mathrm{p}$ (bottom) orbitals to the electronic states.
    }
    \label{fig:bs_all}
\end{figure*}

To visualize the gate-induced shifting of the bands as well as the orbital character of the involved bands, we plot the band structure of the FGT/Gr heterostructure for electric fields $E_\mathrm{z}=\pm10$\,V/nm in Fig.~\ref{fig:bs_all}. The shift of the Dirac cone around K is clearly visible. Furthermore, we observe that the bands closer to $\Gamma$ which contribute strongly to the damping (compare Fig.~\ref{fig:kspace}) are almost entirely made up from Fe-$d$ orbitals with some minor contributions from Te$-p$ orbitals. Contributions from Ge are negligible in this energy range.

\section{Heterostructure results}
\label{app:table}
Complementing the graphical presentation in Fig.~\ref{fig:DFT}, we list the numerical results for the FGT/Gr heterostructure in table~\ref{tab:field_alpha} below. For the evaluation of the polar damping according to the BFS moel, we set the lifetime of electron-hole pairs to $\tau=1$\,ps and temperature $T=10$\,K.
\begin{table*}[htbp]
    \centering
    \caption{Magnetic damping ($\alpha_{\theta\theta}$) as a function of the applied electric field ($E_z$).}
    \label{tab:field_alpha}
    \begin{tabular*}{\textwidth}{l @{\extracolsep{\fill}} *{11}{c}}
        \toprule
        $E_z$\,[V/nm] & -10 & -7 & -5 & -2 & -1 & 0 & 1 & 2 & 5 & 7 & 10 \\
        \midrule
        $\alpha_{\theta\theta}$\,[1] & 0.197 & 0.063 & 0.039 & 0.051 & 0.036 & 0.029 & 0.011 & 0.0099 & 0.0009 & 0.0056 & 0.0063 \\
        \bottomrule
    \end{tabular*}
\end{table*}

\end{document}